\documentclass[onecolumn,amssymb,aps]{revtex4}

\newcommand{\rx}[1]{\ensuremath{\lceil \mathtt{#1} \rfloor}}
\newcommand{\defrx}[2]{\ensuremath{\mathcal{#1} = \lceil \mathtt{#2} \rfloor}}
\newcommand{\defsrx}[3]{\ensuremath{\mathcal{#1}_{#2} = \lceil \mathtt{#3} \rfloor}}
\newcommand{\code}[1]{\ensuremath{\lceil #1 \rfloor}}

\newcommand{\ini}{\ensuremath{\mathcal{I}}}
\newcommand{\aut}{\ensuremath{\mathcal{A}}}
\newcommand{\add}{\ensuremath{\mathcal{L}}}
\newcommand{\autblock}{\ensuremath{\mathcal{B}}}
\newcommand{\autline}{\ensuremath{\mathcal{B}_s}}

  \makeatletter \global\@specialpagefalse \def\@oddhead{{\textit{Constans, P.}}
\hfill \textit{Author field extraction } \thepage} \let\@evenhead\@oddhead
\makeatother

\begin{document}

\title{A Simple Extraction Procedure for Bibliographical Author Field}

\author{Pere Constans}
  \email{constans@molspaces.com} \affiliation{Banyoles, February 2009}

\begin{abstract}
A procedure for bibliographic author metadata extraction from scholarly texts
is presented. The author segments are identified based on capitalization and
line break patterns. Two main author layout templates, which can retrieve from
a varied set of title pages, are provided. Additionally, several disambiguating
rules are described.
\end{abstract}
\maketitle

\section{Introduction}

  Several recognition methods that target bibliographical metadata extraction
from scholarly texts had been described. Based on the underlying methodological
basis, they are formally grouped into two categories. One, named knowledge
representation and template mining, uses prior pattern knowledge to segment
texts and to retrieve data \cite{Ding99p47, Jeng99p99, Giuffrida00p77,
Bergmark00, Kim03p47, Rath04p389, Day07p152}. The other, which is based on
general machine learning techniques, applies statistical devices, concretely
hidden Markov models \cite{Geng04p193, Takasu05p1298, Hetzner08p280}, support
vector machines \cite{Han03p27}, and conditional random fields
\cite{Peng04p329}, to infer segmentations. The knowledge of structural patterns
is thus automatically embedded into transition and conditional probability
matrices that are precomputed from sample sets.

  Practical implementations of these methods, not only the knowledge
representation ones, but to some extend the machine learning ones, include a
priori information of the textual patterns. Described patterns are font size
and typeface variations, HTML markup and delimiters, relative length of title
versus author segments, predefined segment ordering, and textual punctuation
marks. Capitalization and line breaks, albeit being noticed in references
\cite{Bergmark00, Han03p27, Peng04p329}, have eluded much of the attention and
their potential has not been fully exploited.

  This article focuses on the extraction of the author's field. The presented
procedure is based on the identification of simple, yet general templates or
regular expressions. Author segments are recognized solely upon capitalization
patterns and line break delimiters. Section \ref{encoding} describes a
minimalist text encoding or tokenization, and gives the corresponding author
templates. Section \ref{layouts} analyzes layout templates. Both author and
layout templates emerge from the examination of plain text conversions of 2350
title pages. Interestingly, scholarly text layouts constrain author segments to
only two main templates. This is in sharp contrast with general proper name
grammars, which consist of over a hundred rules \cite{Wakao96p418}. Section
\ref{adparticles} provides several disambiguating rules, and describes the use
of a common name prefix lexicon that particularizes the procedure to specific
domains. Section \ref{remarks} gives the details of the data set and other
implementation details.

\section{Text Encoding and Author Name Templates}
  \label{encoding}

  Words, which are defined here as contiguous sequences of two or more letters,
are mapped to a single symbol code. Different symbols are assigned based on
whether all letters are uppercase, only the first letter is uppercase, or at
least the first letter is lowercase. Additionally, initials, line breaks,
relevant punctuation marks, and some special, high-frequency words, have a
particular code. Spaces are only meaningful to discern word boundaries and are
omitted in the encoded string. The complete code mapping is listed in Table
\ref{code_table}.

  This encoding or tokenization, besides simplifying string matching,
highlights the author patterns. The text
\begin{center}
\begin{quotation}
\begin{small}
%\textit{Isaac Newton, Philosophi{\ae{}} Naturalis Principia Mathematica.}
\textit{Philosophi{\ae{}} Naturalis Principia Mathematica\\Isaac Newton}
\end{small}
\end{quotation}
\end{center}
  for example, produces the encoded string \code{LnnnnLnnL}, from which segment
\code{nn} is identified as an author pattern. The inspection of the complete
title page data set produces the following author name templates
\begin{equation}
\defsrx{A}{l}{(?:n\ini n|n\ini \ini n|\ini n|\ini nn|\ini \ini n|\ini \ini \ini n|n\ini nn|nn\ini n|\ini \ini nn|nn|nnn)},
\end{equation}
  and
\begin{equation}
\defsrx{A}{u}{(?:[nN]\ini N|[nN]\ini \ini N|\ini N|\ini [nN]N|\ini \ini N|\ini \ini \ini N|[nN]\ini [nN]N|[nN]N\ini N|\ini
\ini [nN]N|[nN]N|[nN][nN]N)},
\end{equation}
  with initial \ini\ being
\begin{equation}
\defrx{I}{Ip\{0,1\}}.
\end{equation}
  Exceptions to these patterns are one single-letter last name, and two
four-word names on approximately ten thousand names. The relatively frequent
cases of hyphenated names, and names with personal articles, prepositions and
qualifiers, do match the patterns by simply reclassifying hyphen as being
letter, and annexing personal particles in a preprocessing step. While reversed
order naming is frequent in references and databases, only natural ordering
appears in the title page data set.

\begin{table}
\begin{ruledtabular}
\begin{tabular}{cllp{2in}}
Code & Description & \\ \hline
N & Name & \textit{All uppercase} \\
n & Name & \textit{First letter uppercase} \\ 
I & Initial & \textit{Single, uppercase letter} \\
w & Word & \textit{At least first letter lowercase} \\
p & Period & \\
, & Comma & \\
; & Semicolon & \\
: & Colon & \\
\& & And & \textit{Conjunction 'and'} \\
L & Line & \textit{Line break} \\
a & Adparticle & \textit{Special particles that precede words} \\ \hline
o & Other & \textit{Any other symbol except space}
\end{tabular}
\end{ruledtabular}
\caption{Coding scheme.
Word boundaries are identified by non-letter symbols and spaces. Spaces are omitted in the encoded string.}
\label{code_table}
\end{table}

\section{Author Layout Templates}
  \label{layouts}

  Once the text content is extracted from a document, any stylistic layout,
except capitalization and line breaks, is stripped. Then emerges the bare
author layout. Two layouts are found, the single block of authors
\begin{equation}
\defsrx{B}{s}{L\aut([,;\&L]+\aut)*(?=L)},
\end{equation}
  and the multiple block, with the author lines \autline\ being
%\begin{equation}
%\defrx{B}{L\aut([,;\&L]+\aut)*(?=L)}
%\end{equation}
  followed, possibly, by the author addresses
\begin{equation}
\defrx{L}{(L[\hat{\;}L]*)}.
\end{equation}
  The generic author template \aut\ stands for $\aut_l$ and $\aut_u$. The two
are applied sequentially. A three-block layout, for instance, will be
\begin{equation}
\defsrx{B}{3}{(\autline)\add\{0,7\}(\autline)\add\{0,7\}(\autline)}.
\end{equation}

  The set of author \aut\ and layout \autblock\ templates is sufficient to
correctly extract an 83\% of the authors in the data set. To extract the
remaining, the patterns \rx{LnnL} and \rx{LnnnL}, and the corresponding
all-uppercase cases, must be disambiguated.

\section{Adparticles and Lexicons}
  \label{adparticles}

  Most of the \rx{LnnL} and \rx{LnnnL} patterns in the text that are not actual
proper names are parts of uppercase titles spanning multiple lines. Others are
publisher's tags, such as \textit{Open Access}, and a few of them are
addresses. An analysis of over one hundred thousand titles has identified the
most frequent words as being \textit{of}, \textit{the}, \textit{and},
\textit{in}, \textit{for}, \textit{from}, \textit{with}, \textit{to}, and
\textit{on}. On average, preposition \textit{of} appears on any title, and
\textit{on} in one of every ten titles. These high-frequency words are encoded
as \rx{a}, and are referred here as \textit{adparticles}, meaning that they are
connected to, or connect other words. Lexically, adparticles would be
adpositions, adverbs, articles, conjunctions, coverbs, prepositions, or some
verbal forms such as \textit{using}. Adparticles possess, therefore, the
desirable property of permitting to safely scape subsequent words. Adparticle
scape templates are listed in Table \ref{scape_table}, together with a case
example.

  Additionally, a common name lexicon has been recompiled to disambiguate cases
as the previously mentioned \textit{Open Access}, or noisy \textit{Email
Alerts}, which might appear when texts are extracted from abstract web pages.
The lexicon is composed of frequent prefixes, and it is used to lowercase words
before encoding the text. The prefixes are obtained according to the procedure
described in Section \ref{remarks}, in order to avoid lowercasing actual proper
names. Effective lexicons are domain specific, as the word frequency is.
Approximately, fifty prefixes have been sufficient to correctly extract all the
authors from the 2350 item data set.

\begin{table}
\begin{ruledtabular}
\begin{tabular}{llp{2in}}
Example & Template \\ \hline

\textit{Nonlinear Theory of} & \defsrx{S}{1}{aL+[nN]\{1,2\}} \\
\textit{Shallow Shells} & \\
 & \\

\textit{... in Linear and} & \defsrx{S}{2}{a[nNw]\&L+[nN]\{1,2\}} \\
\textit{Sublinear Time} & \\
 & \\

\textit{... and Potentials:} & \defsrx{S}{3}{:L+[nN]\{1,2\}} \\
\textit{Concept Elaboration} & \\
 & \\

\textit{Unrestricted Hartree-Fock Then and} & \defsrx{S}{4}{[nN]*\&L[nN]L} \\
\textit{Now} & \\

\end{tabular}
\end{ruledtabular}
\caption{Scape templates.
Examples of ambiguous cases of capitalized common names, and possible templates to safely scape them.}
\label{scape_table}
\end{table}

\section{Remarks}
  \label{remarks}

\begin{description}

\item[Data Set.]
Author name and layout templates have been identified based on the title page
of 2350 scientific works. They include a variety of journals and publishers, in
addition to self-published drafts and preprints. Most of these works have a
single-block author layout. Approximately 400 works have a multi-block layout.
The set is a curation from an initial set of 2600 works. Exclusion was due to
either an extremely poor conversion to plain text, or due to the existence of
infrequent patterns that would conflict with general templates. Disregarded
patterns are single-letter last names, four-word names, lack of separator among
coauthors, and line breaks between fore and last name.

\item[Common Name Lexicon.]
To lowercase common words without conflicting with real proper names the
following procedure has been devised. Given a list of author names, and a list
of common word candidates, the shortest prefix of each candidate that is not an
author prefix is recorded. In this way, a list of unique shortest prefixes is
obtained, together with the cumulative frequency of each. The most frequent
prefixes are then included in the lexicon. Note that the words \textit{low} or
\textit{water}, even though they are frequent in chemical texts as common
names, are also proper names, and, therefore, are not included. The overlap
between common and proper names might be huge in an international setting.
Still, frequent scientific terms, such as \textit{functional} or
\textit{tetrahedron} appear apart from proper names.

\item[Additional Data and Software.]
Conversions of the title pages to plain text has been accomplished using
\textsc{Xpdf} \cite{Xpdf}. Its \texttt{pdftotext} utility has been modified,
setting the rasterization parameter \texttt{maxIntraLineDelta} to 0.2 in order
to eliminate possible author superscripts.
Approximately one hundred thousand PubMed citations \cite{PubMed} has been
processed to analyze title words and author names. Two lists, one with over
sixty thousand unique words and their frequencies, and another with over one
hundred thousand unique fore and last names, have been recompiled. These two
lists have been used to build the prefix lexicon, as described above. While no
more than fifty prefixes are required to correctly retrieve all the authors
from the data set, for the sake of a greater generality, the lexicon has been
enlarged up to 450 entries in the current implementation.
The procedure for the author field extraction has been implemented in the
\textsc{cb2Bib} program \cite{Constans09}, version 1.1.1, and it is part of its
set of recognition algorithms.

\end{description}

\section{Acknowledgment}

  I am grateful to S. Vega for the careful reading of this manuscript.

\bibliographystyle{unsrt}
\bibliography{algorithms}

\end{document}